# New Magicity of Light Nuclei

C. Samanta[*] and S. Adhikari[*]

*Physics Department, Virginia Commonwealth University, Richmond, Virginia 23284-2000*

A new mass formula capable of explaining the binding energies of almost all the known isotopes from Li to Bi is prescribed. In addition to identifying the new magic number at neutron number N=16 (Z=7-9), pseudo-magic numbers at N= 14 (Z= 7-10), Z=14 (N=13-19), and at N=6 (Z=3-8), the formula accounts for the loss of magicity for nuclei with N=8 (Z=4) and N=20 (Z=12-17). The redefinition of the neutron drip line resulting from this formula further allows us to predict the existence of $^{26}$O, $^{31}$F, $^{32}$Ne, $^{35}$Na, $^{38}$Mg, $^{41}$Al as bound nuclei and $^{28}$O as unbound.

The origin of the unusual stability of nuclei with nucleon numbers 2, 8, 20, 28, 50, 82 and 126, commonly referred to as "magic numbers", was explained more than half a century ago to be due to nuclear shell structure [1-3]. The discovery in 1984 that similar magic numbers also appear in small clusters of Na-atoms [4] was at first surprising since the nature of force holding atomic clusters and nuclei are fundamentally different. In analogy with the stability of magic nuclei, the magic numbers in Na-clusters were explained due to electronic shell closure. However, due to the different nature of the potentials in atomic and nuclear domain, the two sets of magic numbers do not exactly coincide, namely, the magic numbers in the Na-clusters are 2, 8, 20, 40, 58, 92…[4,5] whereas, the nuclear magic numbers are 2, 8, 20, 28, 50, 82...[3].

Nevertheless, these two unlikely fields have found many commonalities such as, giant dipole resonance (which was well known in nuclear physics) is now found to exist in free electron atomic cluster [5]. The root-mean-square (rms) radii of magic nuclei are relatively small compared to other nearby nuclei. Similarly, the electron density distribution in magic atomic clusters is more localized than those that are less stable.

Recently, atomic clusters have been found whose stabilities are not ruled by the electronic shell closure. For example, Na$_8$Mg cluster [6] which contains 10 valence electrons should not be magic, but it is. Similarly, Al$^-_{13}$ and Al$_{13}$K [7], which contains 40 electrons each, are magic while Al$_{13}$Cu which also contains 40 electrons is not [8]. The same phenomena also appear in nuclei. Recently the neutron number N=16 has been found to be a new magic number in only a few neutron-rich nuclei [9]. The central question then arises: why do nuclei and clusters deviate from the magic numbers prescribed by the shell structure and is there a convenient way to predict the occurrence of the anomalous magic clusters /nuclei. While such analysis may be easier in atomic clusters because of the interaction being well known, similar analysis in nuclei poses some difficulties.

There exists in literature mass formulae [3,10-12] to examine the stability of nuclei with varying neutron and proton numbers. However, these formulations are more suitable for medium to heavy mass nuclei. No mass formula so far exists that can account for the exotic properties of the light nuclei near the dripline. We have critically examined the available experimental data [13] from Li to Bi, which has enabled us to prescribe a mass formula capable of explaining the binding energies of almost all the known isotopes from Li to Bi. From a comparison with the experimental one-neutron ($S_n$) or, one-proton ($S_p$) separation energies, we can identify the locations of new magicity or, loss of it. The redefinition of the dripline resulting from this formula further allows us to predict the existence of several bound nuclei beyond the conventional neutron drip line.

To formulate this mass formula we started with the Bethe-Weizsacker (B-W) mass formula [3], which gives a reasonable description of the binding energy of medium to heavy mass nuclei. The B-W formula for the binding energy of a nucleus of atomic number A and proton number Z is,

$$B(A, Z) = a_v A - a_s A^{2/3} - a_c Z(Z-1)/A^{1/3} - a_{sym}(A-2Z)^2/A + \delta$$

(1)

where, $a_v$ =15.85 MeV, $a_s$ =18.34 MeV, $a_c$ =0.71 MeV and $a_{sym}$ =23.21 MeV. The pairing energy term $\delta = +a_p A^{-1/2}$ for even N-even Z, $-a_p A^{-1/2}$ for odd N-odd Z, and 0 for odd A nuclei and, $a_p$ =12 MeV. This mass formula not only underestimates the binding energies of the light nuclei, it also predicts too large a pairing energy for such nuclei (Fig.1, 2).

It is pertinent to note that a mere survey of the one-neutron separation energies of the available nuclei reveals extra stability of certain nuclei. This led to the discovery of a new magic number at N=16 [9] but, the exact Z region could not be clearly established as the corresponding cross section data yielded ambigious results. An anomalous behaviour in the binding energy of very neutron-rich Na isotopes around N=20 was also observed and it was attributed to a loss of magicity [9,14]. Since the B-W mass formula is inadequate in the low mass region, it cannot be used to identify such deviations. Most importantly, a clear evidence of particle stability in $^{31}$F (as opposed to instability in $^{30}$F) was found [15], but according to the B-W formula both $^{30,31}$F should be unstable. It is thus imperative to formulate a mass equation which can help to identify the new magicity or, its loss and, give a proper limit of the neutron-drip line.

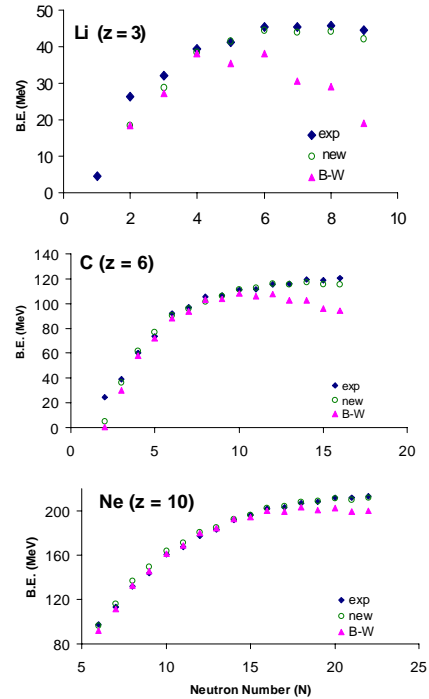

**Fig 1 Binding Energy vs. N, calculated with B-W and new formula**

A careful study reveals that a marked deviation between the experimental binding energy and the prediction of the B-W mass formula occurs in light nuclei, specially when the neutron proton asymmetry is large. This indicates a more complicated dependence of nuclear binding energy on the neutron-proton number arising from a major change in shape and size (like, halo/skin) of the nucleus near the neutron dripline [16]. Recently, near $N \approx Z$ the role of the neutron-proton interaction and its consequences for p-n pairing has been investigated [17]. A steep decrease of the isoscalar p-n pairing energy was suggested with increasing $|N-Z|$. A detailed and systematic search carried out to fit all the isotopes from Li to Bi also supports that and, leads to an additional term in the existing mass formula and a redefinition of the pairing term. The binding energy of any nucleus of mass A is then defined as,

$$B(A, Z)_{new} = a_v A - a_s A^{2/3} - a_c Z(Z-1)/A^{1/3} - a_{sym}(A-2Z)^2/A + \Delta(N,Z) + \delta_{new},$$
(2)

where, $\Delta(N, Z) = |N - 4/3\, Z| \, N^k Z \, e^{-z/3}$, $\delta_{new} = (1 - e^{-z/c})\, \delta$, $k = 0.45$, and $c = 6.0/\text{Ln } 2$, (other constants remaining the same).

While almost all the elements from Li to Bi can be explained by this unique equation (2), binding energies of some very neutron deficient light nuclei are slightly underestimated (Fig.1).

The power of this mass formula is its uniqueness in explaining not only the binding energies but also the one- or, two- neutron and proton- separation energies. Fig. 2 shows a comparison of the $S_n$ data and theoretical calculations for Z= 9 and 10. As the latter does not incorporate the nuclear shell effect, the experimental separation energies near the magic number show distinct deviation from the calculations.

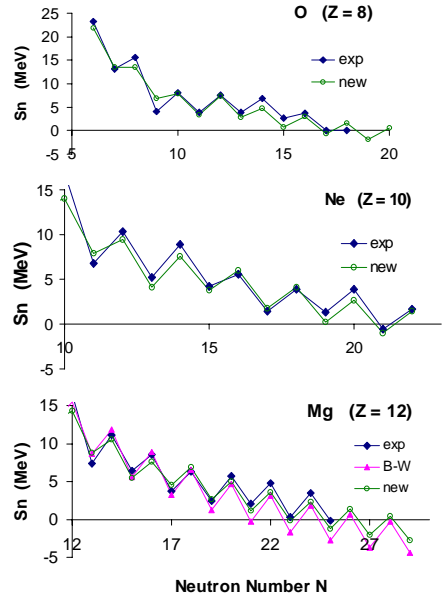

**Fig 2 One - neutron separation energy vs. N and calculation with formula**

Just prior to the magic number, the experimental $S_n$ values are significantly higher and, right after it; the $S_n$ suddenly drops to a lower value. Sometimes, the $S_n$ value does not drop drastically after being high and, we call that particular nucleon number as pseudo-magic, as it reflects pseudo shell closure due to rearrangements of shells that changes both position and width of the shell gaps. In Fig. 2 the known magic number at N=20 (Z=10) can be clearly identified. The magicity at N=20 is found to disappear in the Z= 12 -17 region. Magicity at N=8 also disappears at Z=4 (Fig. 3).

The $S_n$ values from the new formula predicts $^{26}$O and $^{28}$O as bound nuclei with very small binding energies (Fig. 2). However, the two-neutron separation energy($S_{2n}$) is negative for and experiments [15,18] put upper limits to their cross sections to ~ 0.7pb and 0.2pb respectively.

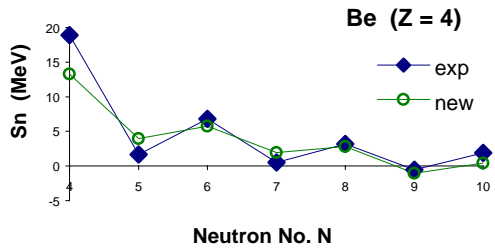

**Figure 3 One-neutron separation energy vs. N, and calculation with new formula**

$^{28}$O, making it unbound (Fig. 4). So far these isotopes of Oxygen are not found

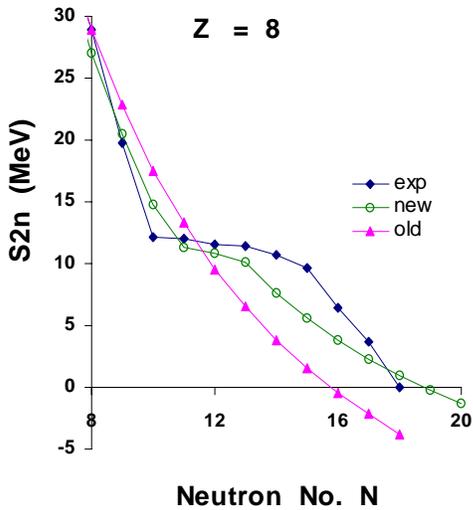

**Figure 4 Two-neutron separation energy vs. N and calculations with old and new formulae**

Absence of the doubly magic $^{28}$O (Z=8, N=20) nucleus is an interesting phenomenon as it indicates a significant change in nuclear structure near the neutron drip line. Infact, we find that in Nitrogen, Oxygen and Fluorine the magicity has moved down to N=16 from N=20. Thus, the new formula helps to resolve the previous ambiguity and ascertains magicity at N=16 for Z= 7- 9. This is supported by the cross section data also [9].

For Z= 7-10, one neutron separation energy is found to be large at N=14 (Fig.2) and, one proton separation energy is large at Z=14 for N=13-19 (Fig.5). In the above region, the nucleon

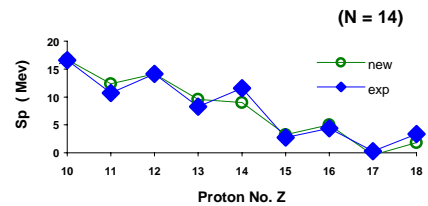

**Figure 5 One-proton separation energy vs. Z and calculation with new formula**

number N=14 can be called as, pseudo-magic. The magicity at N=16 was

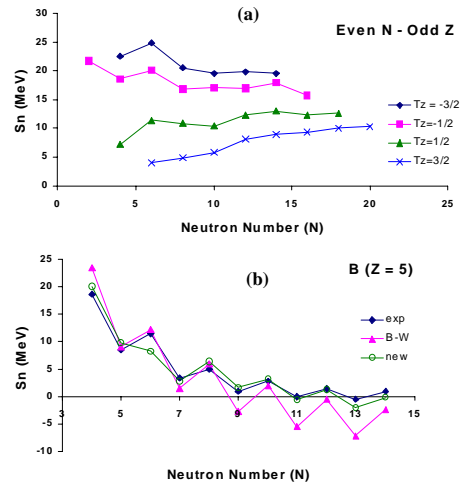

**Figure 6 One-neutrn separation energy (exp) vs. N; (a) for different Tz and (b) calculations with new formula**

predicted to arise from the lowering of

the $2S_{1/2}$ state below the $1d_{3/2}$ [9]. But, the pseudo-magicity at N=14 can occur if the $1d_{5/2}$ remains below $2S_{1/2}$ and the spin-orbit splitting for the d-state is large.

A region of extra stability is found at N= 6 for proton numbers Z= 3-9. The plot of $S_n$ vs. N for different isospin $T_z$ (Fig. 6) shows a clear rise in $S_n$ value for neutron number N=6. The r.m.s. matter radii [19] for all the nuclei ($^9$Li, $^{10}$Be, $^{11}$B, $^{12}$C, $^{13}$N, $^{14}$O) show a distinct bunching at a value ~2.3 fm (Fig. 7) indicating their tight binding. The nucleus $^{15}$F is proton unstable.

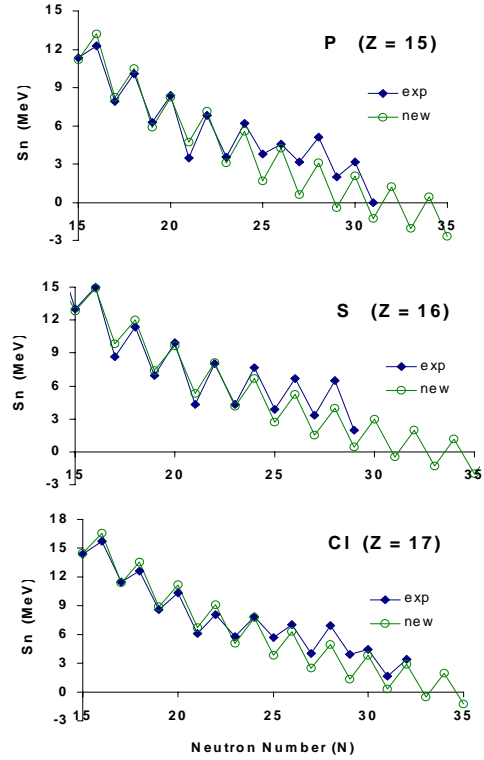

**Figure 8 One-neutron separation energy vs. N, and calculations with old and new formula**

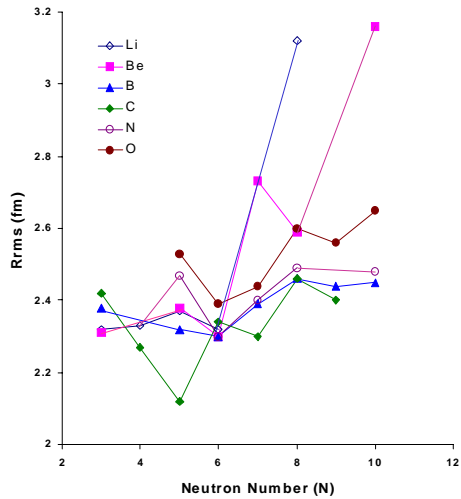

**Figure 7 r.m.s. matter radii vs. N for Z = 3-8**

Another interesting finding is the weakening of the N=28 shell closure in Z=15-17 region (Fig. 7). According to Shell model and relativistic mean field calculations, this arises from deformed prolate ground state associated with shape coexistence [18].

In atomic mixed clusters, the heteroatom at the center was found to cause a narrow depression (or, a local maximum, depending on the electron density) near the center of the potential. This results in unequal shifts in single particle levels. The s-orbital is more influenced than the p and d orbitals as the repulsive centrifugal potentials insulates the higher angular momenta from the narrow central region. Movement of single particle levels causes appearance of new magic numbers and disappearance of the old ones [5]. Similarly, in nuclei away from the β-stability line, the rearrangement of the nuclear shells causes drifting of nuclear magic numbers from the known to unexpected values. This has been already observed in $^{11}$Be for which the ground state spin parity is ½$^+$ instead of ½$^-$ indicating lowering of the $2s_{1/2}$ state below $1p_{1/2}$ [20].

Unlike the B-W formula, the new formula supports experimentally established stability of $^{31}$F (and instability of $^{30}$F) [15]. A survey of $S_n$ and $S_{2n}$ suggests $^{32}$Ne, $^{35}$Na, $^{38}$Mg, and $^{41}$Al as bound nuclei which are beyond the neutron-drip line predicted by the B-W formula.

In summary a "new" mass formula capable of explaining binding energies of almost all the known isotopes from Li to Bi is prescribed. The calculated binding energies, as well as the one-neutron separation energies, are compared with the experimental data and the predictions of the Bethe-Weizsacker mass formula. In addition to identifying the new magic number at N=16 (Z=7-9), we suggest pseudo-magic numbers at N=6 (Z=3-8), N=14 (Z=7-10), and Z=14 (N=13-19). The new formula also accounts for the loss of magicity for nuclei with N=8 (Z=4) and N=20 (Z=12-17). The redefinition of the neutron drip line resulting from this formula further allows us to predict the existence of $^{26}$O, $^{31}$F, $^{32}$Ne, $^{35}$Na, $^{38}$Mg, $^{41}$Al as bound nuclei which are beyond the conventional (B-W) neutron dripline. It is now left as a challenge to find more fundamental basis for this mass formula and its origin based upon the basic nucleon-nucleon interaction.

We would like to thank P.Jena for stimulating discussions on atomic clusters and, I. Samanta for his help in computation.

## References:

[1] M.G.Mayer, Phys. Rev. **75**, 1969 (1949); ibid, Phys. Rev. **78**, 16 (1950)

[2] D.Haxel, J.H.D.Jensen and H.E.Suess, Phys. Rev. **75**, 1766 (1949); ibid. Z. Physik **128**, 295 (1950)

[3] K. Heyde, *"Basic ideas and concepts in nuclear physics"*, IOP publication, 1999

[4] W.D.Knight et al., Phys. Rev. Lett., **52**, 2141 (1984)

[5] C.Yannouleas, P.Jena and S.N.Khanna, Phys. Rev. **B 46**, 9751 (1992)

[6] M.M.Kappes et al., Phys. Lett. **119**, 11 (1985)

[7] B.K.Rao, S.N.Khanna and P.Jena, Phys. Rev. **B 62**, 4666 (2000)

[8] O.C.Thomas, W.Zheng and K.H.Bowen, J. Chem. Phys. (in press); S.N.Khanna et al., J.Chem. Phys. (in press)

[9] A.Ozawa et al., Phys. Rev. Lett. **84**, 5493 (2000)

[10] W. D. Myers and W. J. Swiatecki, Nucl. Phys. **81**, 1 (1966)

[11] R.C.Nayak and L.Satpathy, Atomic data and Nucl. Data tables **73,** 213 (1999)**,** L.Sathpathy and R.C.Nayak, J. Phys. G: Nucl. Part. Phys. **24**, 1527 (1998); ibid. Phys. Rev. Lett. **51**, 1243 (1983)

[12] S. Liran, A. Marinov, and N. Zeldes, Phys. Rev. **C62**, 047301 (2000)

[13] Nuclear wallet cards, January 2000; Brookhaven National Laboratory

[14] C. Thibault et al., Phys. Rev. **C12**, 644 (1975)

[15] H. Sakurai et al., Phys. Lett. **B 448**, 180 (1999)

[16] I. Tanihata, J. Phys. G: Nucl. Part. Phys.**22**, 157 (1996)

[17] G.Ropke et al., Phys. Rev. **C 61**, 024306 (2000)

[18] Z.Dlouhy et al, *Proc. of the 9$^{th}$ Int. Conf. On Nuclear Reaction Mechanisms, Ed. Gadioli, Varena, June 5-9, 2000; see references therein.*

[19] A.Ozawa et al., Nucl. Phys. **A 608**, 63 (1996); see references there in

[20] M.Fukuda et al., Phys. Lett. **B 268**, 339 (1991)